\documentclass[]{spie}  %>>> use for US letter paper
\usepackage{amsmath,amsfonts,amssymb}
\usepackage{graphicx}
\usepackage{hhline}
\usepackage[colorlinks=true, allcolors=blue]{hyperref}

\title{SAM3D: Zero-Shot Semi-Automatic Segmentation in 3D Medical Images with the Segment Anything Model}

\author[a]{Trevor J. Chan*}
\author[a]{Aarush Sahni*}
\author[a]{Yijin Fang*}
\author[a]{Jie Li*}
\author[a]{Alisha Luthra*}
\author[a,b]{Alison Pouch}
\author[b,c]{Chamith S. Rajapakse}

\affil[a]{Department of Bioengineering, University of Pennsylvania, Philadelphia, USA}
\affil[b]{Department of Radiology, University of Pennsylvania, Philadelphia, USA}
\affil[c]{Department of Orthopaedic Surgery, University of Pennsylvania, Philadelphia, USA}

\authorinfo{Further author information: (Send correspondence to T.J.C.: tjchan@seas.upenn.edu)}

\begin{document} 
\maketitle
\def\thefootnote{*}\footnotetext{These authors contributed equally}\def\thefootnote{\arabic{footnote}}

\begin{abstract}
We introduce SAM3D, a new approach to semi-automatic zero-shot segmentation of 3D images building on the existing Segment Anything Model. We achieve fast and accurate segmentations in 3D images with a four-step strategy involving: user prompting with 3D polylines, volume slicing along multiple axes, slice-wide inference with a pretrained model, and recomposition and refinement in 3D. We evaluated SAM3D performance qualitatively on an array of imaging modalities and anatomical structures and quantify performance for specific structures in abdominal pelvic CT and brain MRI. Notably, our method achieves good performance with zero model training or finetuning, making it particularly useful for tasks with a scarcity of preexisting labeled data. By enabling users to create 3D segmentations of unseen data quickly and with dramatically reduced manual input, these methods have the potential to aid surgical planning and education, diagnostic imaging, and scientific research.
\end{abstract}


\section{INTRODUCTION}
\label{sec:intro}

Image segmentation is a foundational problem in both medical practice and research. Segmentation plays a critical role in surgical planning and interventional radiology \cite{virzi2020comprehensive,fang2020deep}, it is used to calculate common clinical metrics \cite{kitano2019accuracy}, and it is a component of many currently used and proposed diagnostic tools \cite{chaddad2018deep,lambin2017radiomics}. 

Current automated approaches to image segmentation predominantly use deep learning models trained on vast quantities of labeled data. In medicine, these models typically achieve high performance through hyperfixation: they train on a single anatomical region imaged using a single modality. While narrowly effective, such an approach requires gathering a large amount of annotated data, which can be time consuming and expensive. And, because training datasets are imperfect, these models are also susceptible to brittleness and bias \cite{o2017imaging,lu2016assessing}. Lastly, many medical imaging modalities acquire 3D images, which drastically increases the difficulty of storing, annotating, and processing sufficiently large and diverse datasets.

Because of this, there is a growing interest in semi-automatic approaches for general image segmentation. These methods offer a compelling compromise: by accepting a small decrease in speed and convenience compared to their fully-automated counterparts, we can have both increased reliability and greater generalizability. Early approaches to semi-automatic segmentation include thresholding and region-growing methods for 2D and 3D \cite{kohler1981segmentation}, watershed and active contour methods \cite{beucher1992watershed,kass1988snakes,yushkevich2006user}, and atlas and multi-atlas-based segmentation methods \cite{cabezas2011review,lotjonen2010fast,wang2013multi}. With the advent of deep learning-based segmentation and the release of off-the-shelf models and pipelines \cite{isensee2021nnu,wasserthal2023totalsegmentator}, high-quality automatic segmentation is easier than ever. However, these models still require large amounts of domain-specific training data, so researchers have sought to combine the speed and performance of deep learning methods with the generalizability of previous semi-automatic algorithms \cite{diaz2022deepedit,roy2020squeeze}.

In 2023, Kirillov et al. introduced Segment Anything \cite{kirillov2023segment}, a promptable semi-automatic segmentation model trained on a dataset of over 1 billion masks in a wide array of 2D images. The architecture of the segment anything model (SAM) is simple: it consists of an image encoder, a prompt encoder, and a decoder, which takes the image and prompt embedding and predicts a 2D mask. Prompting, an additional user input to inform the model of which structure to segment, could be supplied one of in four formats. Three of these: points, boxes, masks, describe the location and shape of the object of interest in the image. The fourth, text, describes the semantics of the object. Due in large part to the quantity and diversity of its training data, the segment anything model (SAM) displays remarkable zero-shot segmentation performance: it is able to segment types of images unseen during training.

In this work, we extend the SAM to 3D medical images with a novel prompting, slicing, and recompositing scheme. We test our method on a wide array of 3D medical images and show that it is capable of generating high-quality masks of diverse anatomies across a range of imaging modalities. By dramatically reducing the time and effort required to obtain 3D segmentations on unseen data, these methods have the potential to accelerate both clinical and scientific workflows and improve future fully automatic segmentation tools.

\section{Methods}
\label{sec:methods}
In principle, extending the zero-shot performance of SAM to 3D images is fairly simple. We can slice a 3D image into many 2D images, add appropriate prompting to these images, segment using the pretrained model, and recompose the results into a 3D mask. In practice, multiple decisions regarding the strategy of volume decomposition, prompting, and recomposition affect both the quality of the final 3D mask as well as the amount of human effort and time required. 

When adding prompting to a 3D image, there is a trade off between precision and time; on one hand, prompting every 2D slice individually can yield accurate segmentations but it is very time consuming. On the other, using the same prompt for each 2D slice, such as defining a single 3D bounding box, saves time but introduces ambiguities when inferencing the model. A second major problem occurs in recomposition. Segmenting many individual slices multiplies the likelihood of a bad segmentation prediction, and these errors appear as unrealistic artifacts when images are recomposed in 3D. In the 2D case, these can be easily fixed with revised prompting, but this is not efficient for volumetric images.

\begin{figure}[t]
\centering
\includegraphics[width=\textwidth]{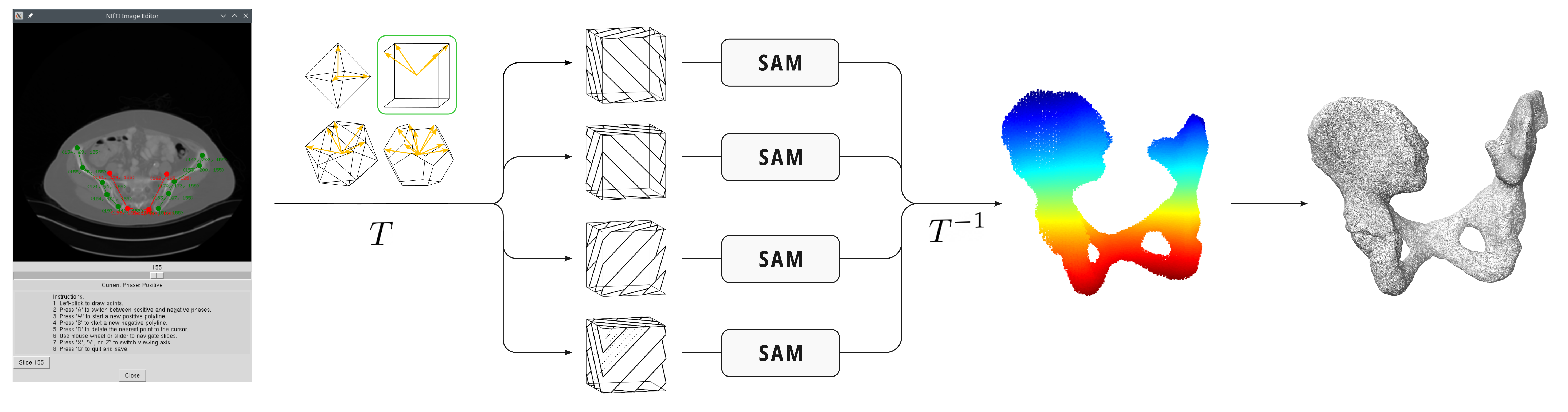}
\caption{An overview of the segmentation method comprising: polyline prompting on a 3D image, slicing along rotationally equispaced axes, 2D inference using SAM, recomposition into a dense point cloud, and voxelization/meshing.} 
\vspace{-8pt}
\label{fig1}
\end{figure}

We address these problems using a novel prompting strategy and a novel slicing strategy. User prompts take the form of polylines (positive and negative) added to slices of a 3D volume. At inference, we calculate the intersections between 1D line segments and the 2D plane we are segmenting to obtain a set of positive and negative prompt points which are then used to condition the pretrained model. In this manner, we can add precise prompting to an entire 3D volume consisting of hundreds of slices while manually prompting fewer than 10 slices.

To address errors that occur during segmentation, we resort to a multi-slicing strategy that produces redundancy in the segmentation outputs. While a typical pipeline for 3D segmentation with a 2D model might traverse along a single axis, we traverse along a predefined set of axes. The resulting planes uniformly cover the 3D volume and, because they intersect, they allow the pretrained segmentation model to see the same spatial location in the volume multiple times, giving it that many chances to segment it correctly. In order to ensure an unbiased distribution of slices, we select orthogonal planes along rotationally equispaced axes. Depending on the complexity of the anatomy being segmented, more or fewer axes may be called for, so the user has the option of using between 3 and 10 independent axes, resulting in a small trade off between accuracy and runtime. And, by taking larger steps along each axis, we can reduce the total number of segmentations the model needs to perform, further reducing runtime.

Recomposing 2D masks back into 3D occurs in two steps. First, we convert all segmentation masks into 2D point clouds and transform the points back into the global 3D reference frame. Second, we filter to remove outlier points in low-density regions. This step exploits the redundancy of the 2D planes: areas in erroneously segmented 2D slices are sparse in 3D space, allowing them to be efficiently removed without affecting the correctly segmented 3D shape. The final point cloud is voxelized and turned into an image mask or an object mesh.

For a typical structure in a $256^3$ voxel image, this entire process takes a few minutes, roughly a third of which is devoted to user prompting and post-processing and two thirds to image transformations and model inference. All experiments and evaluations were performed on a workstation with a single Nvidia 3090 gpu.

\begin{figure}[b]
\includegraphics[width=\textwidth]{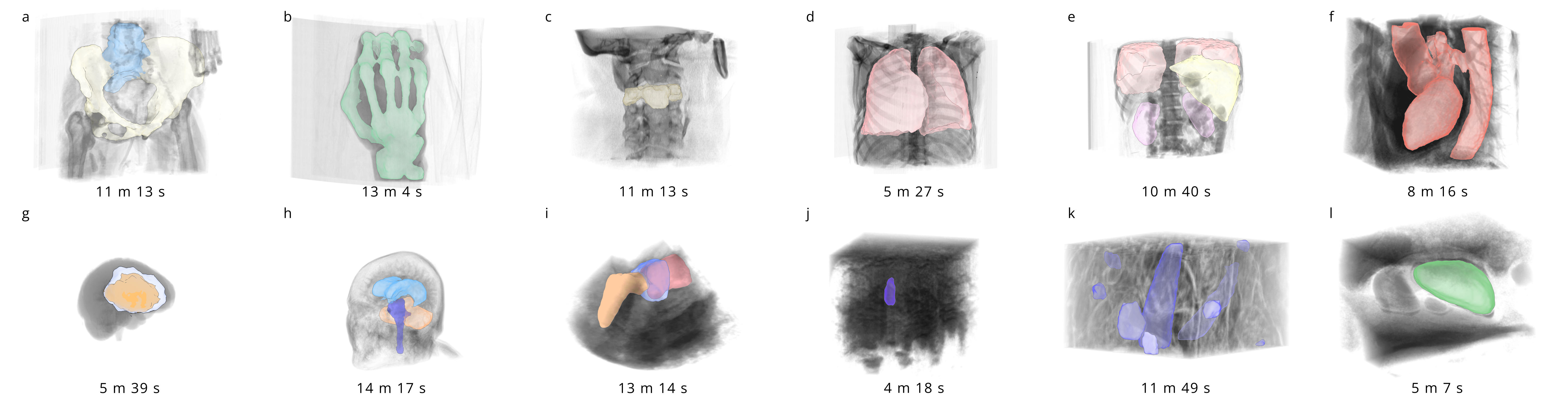}
\caption{Visualizing diverse segmentation performance. (a) Pelvis and sacral spine in CT. (b) Skeleton in \textit{ex vivo} CT. (c) Cervical vertebra 3 in CT\cite{loffler2020vertebral}. (d) Lungs in chest CT\cite{ma2021toward}. (e) Lungs, liver, and kidneys in abdominal CT\cite{Landman2015btcv}. (f) Oxygenated blood pool in CT angiogram. (g) Glioblastoma tumor and edema in FLAIR MRI\cite{menze2014multimodal}. (h) Lateral ventricles, cerebellum, and brain stem in T1 MRI. (i) Left ventricular outflow tract, aortic valve, and aortic root in 3D TEE. (j) Tumor lesion in 3D breast ultrasound\cite{tdscabus}. (k) Hippocampal axonol neurons in volumetric scanning electron microscopy (SEM)\cite{lucchi2011supervoxel}. (l) Tobacco leaf cell central vacuoule in volumetric SEM\cite{wickramanayake2023conventional}.} 
\vspace{-8pt}
\label{fig2}
\end{figure}

To assess the accuracy and generality of our segmentation model, we tested it on a range of image modalities and anatomical structures. We selected datasets and images that capture a wide array of shapes, length scales, and image qualities (Figure \ref{fig2}). For image preprocessing, we resampled the image volumes into isotropic voxel dimensions prior to input into our model. Prompting and postprocessing steps were performed by the authors, and reported times throughout the manuscript represent the duration of the full pipeline, including slicing, prompting, inference, and postprocessing. All data used is either publicly available or was approved for use in this study by an institutional review board.

We further quantified the accuracy of these segmentations using two 3D medical imaging datasets: the Beyond the Cranial Vault (BTCV) dataset for organ segmentation in abdominal CT \cite{Landman2015btcv}, and the Brain Tumor Segmentation (BraTS2021) dataset for glioblastoma in 4 MRI contrasts \cite{menze2014multimodal}. Both datasets have manually labeled masks which serve as ground truth comparisons. In addition, we benchmarked the performance of our method against the popular MedSAM model \cite{ma2024segment} on a range of tasks including brain tumor and metastases segmentation in MRI, pelvic segmentation in CT, and aorta segmentation in transesophageal echocardiography (TEE).

\begin{figure}[t]
\includegraphics[width=\textwidth]{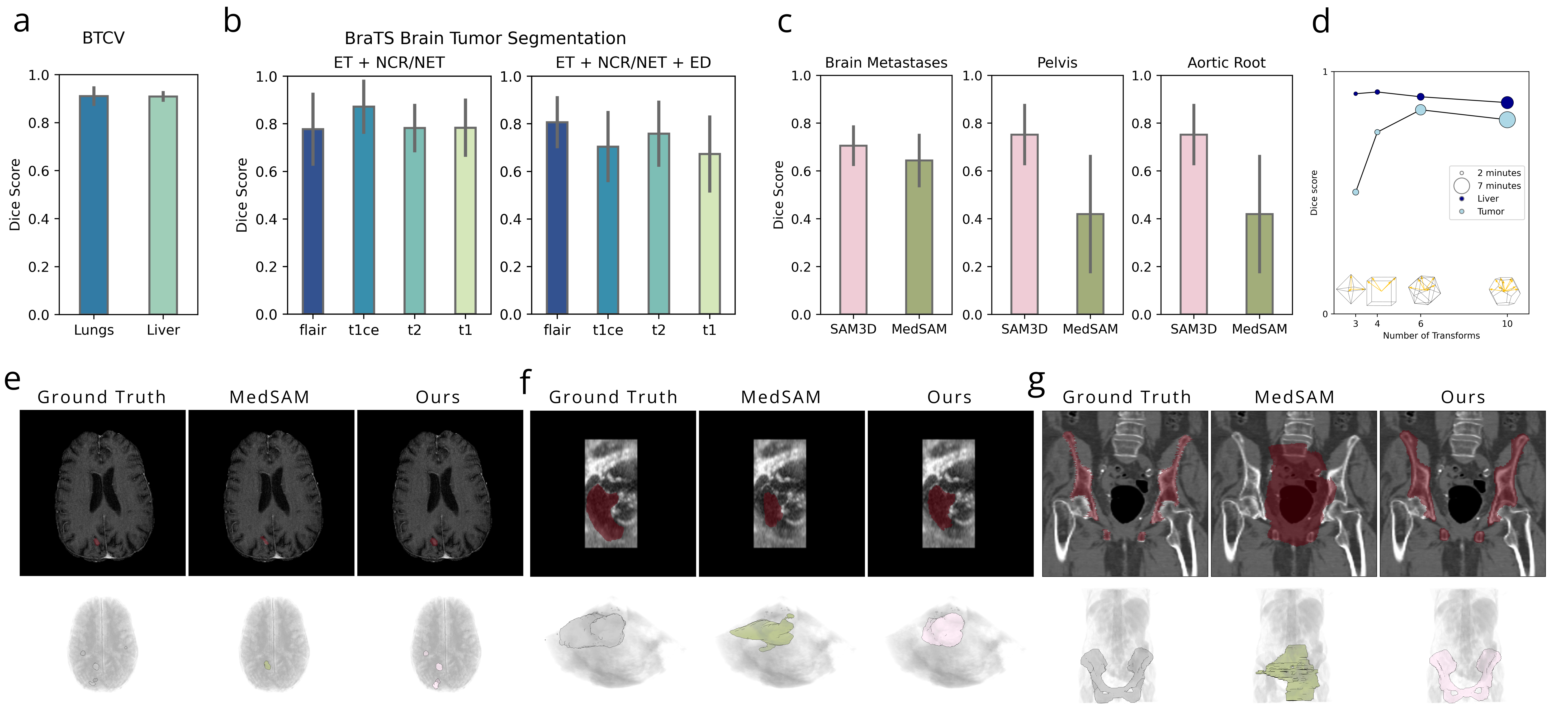}
\caption{Quantification of segmentation accuracy on benchmark datasets, comparison against MedSAM, and additional experiments. (a) Dice score calculated for the lung and liver masks ($n=16$) on the BTCV dataset. (b) Dice score calculated for the tumor region (enhanced + nonenhanced tumor/necrotic) and the tumor+edema regions ($n=8$) for 4 MRI contrasts on the BraTS dataset. (c) Dice scores for segmentation performance on brain metastases ($n=4$), pelvis ($n=3$), and aorta ($n=4$) for both our method (SAM3D) and MedSAM. (d) An analysis of segmentation accuracy and time as a function of the number of transforms chosen for liver segmentation in the BTCV dataset ($n=3$) and tumor segmentation in the BraTS dataset ($n=3$) suggests that the optimal number of transforms to use depends heavily on the anatomical structure to segment. (e-g) Representative segmentation predictions shown in 2D and 3D in three zero-shot tasks for our method and MedSAM.
} \label{fig3}
\end{figure}

\section{Results}
\label{sec:results}

We found that the our model generated high quality masks for a range of imaging modalities and anatomical structures (Figure \ref{fig2}). The time required for each segmentation depended largely on the size and complexity of the anatomy being segmented; intricate structures and images with multiple components took more time to segment, but across the board our method was far faster and easier than slice-wise manual segmentation.

In the BTCV dataset, we compared segmentation performance for the liver and lungs and showed high accuracy for each (Figure \ref{fig3}a). In the BraTS dataset, we performed two segmentations: one for the tumor regions, comprising the enhancing tumor (ET) and necrotic/non-enhancing tumor (NCT/NET), and a second for the tumor region and the surrounding edema (ET+NCT/NET+ED). We found similarly high performance across 4 contrasts (T1, T1-contrast enhanced, T2, FLAIR) (Figure \ref{fig3}b). We also showed consistent mask predictions from the model for scans of the same patient with different MRI contrasts, demonstrating a robustness to pixel-level changes (Figure \ref{fig3}c).

We also compared the zero-shot performance of our model to that of the popular MedSAM model in 3 difficult segmentation tasks. To the best of our knowledge, MedSAM was not explicitly trained on either pelvic segmentation in CT or aortic segmentation in TEE despite having trained on a large variety of data including MRI, CT, and ultrasound. SAM, the model our method uses, was not trained on any medical images. Ground truth masks for each modality were created manually using itk-SNAP. We found that our model achieves significantly higher accuracy than MedSAM in the pelvic segmentation and aortic segmentation tasks (Figure \ref{fig3}c). In the brain tumor segmentation task, for which MedSAM was explicitly trained, the models performed comparably. Our method requires less time on average to perform a segmentation compared to MedSAM. Both models were orders of magnitude faster than manual segmentation, which took on average over an hour to produce a single 3D mask.

To determine the effect of varying the number of transforms used for prompting and slicing, we segmented an additional 3 livers and tumors with 3, 4, 6, and 10 transforms each. We observed that for large, simple structures, such as the liver, 3 or 4 transforms, representing the octahedral and cubic axes, is sufficient. On the other hand, more complex structures, including some brain tumors, required 6+ transforms to reach peak accuracy (Figure \ref{fig3}d). As the amount of time and user input required for a segmentation scales with the number of transforms used, we suggest setting the number of transforms based on the complexity of the anatomy being segmented. 
We report the time required to segment each anatomical structure in the BTCV and BraTS dataset in table \ref{tab1}.

\setlength{\tabcolsep}{14.5pt}
\begin{table}[ht]
\caption{A comparison of segmentation times.}\label{tab1}
\centering
\renewcommand{\arraystretch}{1.0} % Increase the table row spacing
\begin{tabular}{llll}
Method & Dataset & Segmentation Target & Time Taken (Mean +/- Std.) \\
\noalign{\hrule height0.8 pt}\hline
SAM3D & BTCV\cite{Landman2015btcv}  & Liver & 4 min +/- 60 sec \\
SAM3D & BTCV\cite{Landman2015btcv} & Lungs & 3 min 38 sec +/- 51 sec \\
SAM3D & BTCV\cite{Landman2015btcv} & Kidneys & 3 min 50 sec +/- 1 min 47 sec \\ \hline

SAM3D & BraTS\cite{menze2014multimodal} & ET + NCT/NET & 4 min 14 sec +/- 1 min 5 sec \\
SAM3D & BraTS\cite{menze2014multimodal} & ET + NCT /NET+ ED & 7 min 50 sec +/- 28 sec \\\hhline{====}

SAM3D & MRI & Metastases & 3 min 31 sec +/- 1 min 57 sec \\
SAM3D & CT & Pelvis & 6 min 22 sec +/- 1 min 1 sec \\
SAM3D & TEE & Aorta & 4 min 21 sec +/-  42 sec \\ \hline

MedSAM\cite{ma2024segment} & MRI & Metastases & 7 min 40 sec +/- 1 min 17 sec \\
MedSAM\cite{ma2024segment} & CT & Pelvis & 7 min 2 sec +/- 49 sec \\
MedSAM\cite{ma2024segment} & TEE & Aorta & 7 min 3 sec +/-  20 sec \\ \hline

Manual & CT & Pelvis & 1 hour min 28 min +/- 20 min \\
Manual & TEE & Aorta & 1 hour 20 min +/-  44 min \\ \hline
\end{tabular}
\end{table}

\section{Discussion}
\label{sec:discussion}
We introduce a new method for zero-shot semi-automatic 3D image segmentation that leverages a pretrained 2D segmentation model and demonstrate strong results on a variety of 3D images. 

Despite SAM not having seen any 2D or 3D medical imaging data during training, it still performs highly, suggesting a few insights. First, it seems that a learning a true semantic understanding of the data might not be necessary; regional intensities, gradients, and textures, combined with sufficient prompting, are enough to reliably segment a wide variety of structures found in medical and scientific images. (This complements related work in synthetic data for image learning \cite{madhusudana2021revisiting, dey2024anystar}.)

Second, while it may seem unintuitive to use a 2D model to perform 3D segmentation, the approach carries numerous benefits. 2D data is far less expensive to collect and store and far easier to train models on, and this means that the best 2D models, such as SAM, have seen orders of magnitude more data than the best 3D models. For zero-shot segmentation, and possibly many other tasks, the benefit of seeing more out-of-domain data may actually outweigh the benefit of seeing data that is closely aligned with the desired task.

There are some notable limitations of SAM3D. As is the case for many segmentation models, including the base SAM, SAM3D has poor performance when labeling thin and branching structures. A second limitation is that, because the base 2D segmentation model has not been trained on medical images, it lacks any relevant domain knowledge. Multiple groups have taken the base SAM model and fine-tuned it on a medical domain \cite{ma2024segment,huang2024segment,mazurowski2023segment,wu2023medical}, and demonstrate improved results compared to the base model. Inserting these models as drop-in replacements for the SAM in our method could further improve segmentation performance and efficiency.

\section{Conclusion}
\label{sec:conclusion}
SAM3D is a semi-automated, zero-shot 3D segmentation model capable of producing accurate segmentations across a range of structures and images. There are numerous potential uses for the proposed method. In scientific research, semi-automatic segmentation could be used in data-limited regimes or as a means to initially label training datasets. In medicine, 3D segmentation has widespread applications in surgical planning, diagnostic imaging, and radiomics. The use of this tool could save physicians time while mitigating the risks of bias and unpredictability that plague fully automated models. By enabling faster, easier, and more accurate 3D segmentation, we hope that these methods will aid clinicians and researchers, accelerate the creation of large-scale 3D datasets, and spur development in general 3D segmentation models.

% References
\bibliography{report} % bibliography data in report.bib
\bibliographystyle{spiebib} % makes bibtex use spiebib.bst

\end{document}